\documentclass[aps,prd,onecolumn,nofootinbib]{revtex4-2} 
\usepackage{titlesec}
\titleformat*{\section}{\LARGE\bfseries}
\titleformat*{\subsection}{\Large\bfseries}
\titleformat*{\subsubsection}{\large\bfseries}

\usepackage{epsfig}
\usepackage{enumitem}

\usepackage{graphics}     
\usepackage{graphicx}     
\usepackage{subfigure}
\usepackage{longtable}     
\usepackage{url}          
\usepackage{bm}           
\usepackage{natbib}
\usepackage{textpos}
\usepackage{amsfonts,amsmath,amssymb,mathrsfs,bm,amsthm}
\usepackage{float}
\usepackage{booktabs}
\usepackage{array}
\usepackage{tabu}
\usepackage{dcolumn}
\usepackage{rotating}
\usepackage{ulem}

\newcommand{\RN}[1]{%
  \textup{\uppercase\expandafter{\romannumeral#1}}%
}

\usepackage{nicefrac}
\usepackage{amsthm}
\usepackage{color}
\usepackage{cancel}
\usepackage{appendix}
\usepackage{makecell}
\usepackage{upgreek}

\usepackage[colorlinks=true,linkcolor=blue,citecolor=blue,urlcolor=blue,bookmarksopen]{hyperref}

\newcommand{\appropto}{\mathrel{\vcenter{
  \offinterlineskip\halign{\hfil$##$\cr
    \propto\cr\noalign{\kern2pt}\sim\cr\noalign{\kern-2pt}}}}}
\renewcommand{\v}[1]{\boldsymbol{#1}}		

\begin{document}

\title{\Large{
Probing Lorentz-violating scalar forces and dark matter with 
\\precision matter and antimatter experiments
}}

\author{{\large\bf \rule[30pt]{0pt}{0pt}
Yevgeny~V.~Stadnik 
}}


\raggedbottom


\maketitle

\vspace{3mm}
\begin{center}
\Large{\textbf{Abstract}}
\end{center}
\large

We explore a class of Lorentz-violating scalar-field models where the effects of the scalar field on antimatter can be strongly enhanced compared to matter. 
Non-trivial differences in phenomenology arise for spectroscopy measurements depending on whether the scalar field changes slowly or rapidly on the atomic scale. 
We highlight opportunities for precision tests in the laboratory including dark matter searches, in particular with spectroscopy and interferometry measurements, as well as certain types of antimatter experiments.

\vspace{200mm}

\large


\textbf{Introduction} --- 
Various astrophysical and cosmological observations at different epochs indicate that the Universe is predominantly composed of dark matter (DM) and dark energy, with only a mere five percent due to the ordinary matter that makes up stars, planets, dust and interstellar gases \cite{PDG_2026_review}. 
Observations also indicate that there is much more matter in the Universe relative to antimatter than would be expected based on the known sources of CP violation within the framework of standard electroweak baryogenesis \cite{Dine_2003_baryogenesis}. 
Unravelling the identity and microscopic properties of DM, as well as the origins of the matter-antimatter asymmetry, remain among the greatest challenges in modern science. 
Ultralight bosons with sub-eV masses are a compelling candidate to explain the DM \cite{Ferreira_2021_ULDM-review}, as well as other problems outside of cosmology, including the strong CP problem \cite{Kim-Carosi_2010_Axions-review} and electroweak hierarchy problem \cite{Graham_2015_Relaxion}. 
Such DM bosons are expected to produce \textit{wavelike} signatures due to their high number density that are distinct from the \textit{particlelike} signatures of weakly-interacting massive particles (WIMPs) with masses in the $\sim \textrm{GeV} - \textrm{TeV}$ range~\cite{Roszkowski_2018_WIMP-review}. 

Background bosonic fields also feature prominently in models featuring spontaneous Lorentz and CPT violation \cite{Kostelecky_1998_SME}. 
In such models, apparent breaking of the CPT symmetry can manifest in a particular frame of reference due to interactions of the background field that create an asymmetry in matter and antimatter observables. 
For example, a background field can perturb the Larmor frequencies of matter and antimatter particles such that their measured magnetic moments in a given reference frame differ according to $\mu_\textrm{matter} = +\mu (1+\varepsilon)$ and $\mu_\textrm{antimatter} = -\mu (1-\varepsilon)$, where $\varepsilon$ is a small frame-dependent dimensionless correction to the magnetic moment parameter $\mu$. 
In this traditional scenario, the magnitude of the correction to the experimental observable in matter and antimatter systems is the same, but the corrections shift the observable in opposite directions for matter and antimatter. 
Comparisons of observables in matter and antimatter systems, such as hydrogen/antihydrogen atoms \cite{ALPHA_2018_Hbar_expt,ALPHA_2025_Hbar_expt}, protons/antiprotons \cite{BASE_2017_antiproton_MM,BASE_2017_proton_MM,BASE_2022_anti-proton_q2m} and electrons/positrons \cite{Dehmelt_1987_positronium_g-factor,Gabrielse_2023_electron_g-factor}, enable direct and unambiguous tests of the CPT symmetry, though in some cases indirect tests of the CPT symmetry are possible using matter-based systems alone \cite{Safronova_2018_AMO-review}. 

In this paper, we explore a class of Lorentz-violating scalar-field models where the effects of the scalar field on antimatter can be strongly enhanced compared to matter in the laboratory frame. 
We find that in the case of shifts of atomic energy levels due to intra-atomic forces mediated by the exchange of a scalar particle or accelerations exerted on test bodies, effects on non-relativistic antimatter systems can be strongly enhanced compared to that on analogous matter systems. 
On the other hand, in the case when the scalar field changes slowly on the atomic scale (which can occur, e.g., due to a scalar DM field of cosmological origin or an Earth-sourced scalar field acting on atoms in the laboratory), changes in energy intervals in a matter-based atom and its anti-atom counterpart are the same at leading order. 
These qualitatively different regimes lead to a non-trivial phenomenology and, as we highlight in this paper, present opportunities for spectroscopy and interferometry-based searches for ultralight scalar DM, spectroscopy and antimatter freefall experiments searching for new long-range forces, as well as antimatter spectroscopy experiments probing new (sub)atomic-scale forces.

\textbf{Theory} --- 
We consider the following type of operator structure for a scalar field $\phi$ coupling to standard-model (SM) fermions: 
\begin{equation}
\label{LIV_scalar_interaction}
\mathcal{L}_\textrm{int} = - \phi \sum_\psi g_\psi \bar{\psi} \left( 1 - \gamma^0 \right) \psi  \, , 
\end{equation}
where $\psi$ denotes a SM fermion field, $\bar{\psi} = \psi^\dagger \gamma^0$ its Dirac adjoint, $\gamma^0 = \bigl[ \begin{smallmatrix} +I_2 &0 \\ 0 & -I_2 \end{smallmatrix} \bigr]$ is the ``temporal'' Dirac matrix, and $g_\psi$ represent the coupling constants to different fermion species. 
The Dirac matrix structure $1-\gamma^0$ in (\ref{LIV_scalar_interaction}) projects a Dirac spinor $\psi$ onto (twice) its lower component, which in the non-relativistic limit is suppressed by powers of velocity-squared $v^2/c^2 \ll 1$ for matter, while there is no such suppression for antimatter.\footnote{Unless explicitly indicated otherwise, we adopt the natural system of units $\hbar = c =1$, where $c$ is the speed of light in vacuum and $\hbar$ is the reduced Planck constant.} 
The $\bar{\psi} (1-\gamma^0) \psi$ bilinear structure in (\ref{LIV_scalar_interaction}) may arise in the laboratory frame within a manifestly Lorentz covariant theory if Earth moves with respect to some medium or preferred background frame at a non-relativistic speed, similarly to how the presence of a medium apparently breaks the Lorentz symmetry. 
The full operator structure in a Lorentz covariant theory will also contain terms involving the ``spatial'' Dirac matrices $\gamma^i = ( \gamma^1, \gamma^2, \gamma^3 )$, whose magnitude will depend on the specific reference frame. 
The focus of our present work is on spin-independent interactions, where the long-range contribution to the inter-fermion potential scales as $\propto 1/r$ with the inter-fermion separation $r$. 
Therefore, we neglect such $\gamma^i$ terms, which give rise to potential contributions that fall off faster than $1/r$ and generally involve couplings to fermion spin.

In Appendix~\ref{Appendix:A}, we present a model of a complex scalar DM field where a $\bar{\psi} (1-\gamma^0) \psi$ structure may arise in the presence of our Galactic DM halo. 
In this case, the relative speed between the Sun and local DM is $v_\textrm{DM} \approx 220 \, \textrm{km/s}$. 
Another possibility is that the preferred background frame is set by the cosmic microwave background (CMB) reference frame, which the Sun moves relative to at a speed of $v_\textrm{CMB} \approx 370 \, \textrm{km/s}$. 
In either case, it is difficult to imagine a situation where the laboratory frame can sustain a speed relative to the preferred frame of less than Earth's orbital speed around the Sun, $v_\oplus \approx 30 \, \textrm{km/s}$. 
Hence we expect speeds of the laboratory relative to the preferred frame to lie in the non-relativistic range $v_\textrm{bg}/c \sim 10^{-4} - 10^{-3}$. 
For comparison, the typical speeds of electrons and nucleons in an atom are $v_e/c \sim \alpha$ and $v_N/c \sim 0.2$, respectively, where $\alpha \approx 1/137$ is the electromagnetic fine-structure constant. 
This hierarchy in speeds ensures that we can construct observables that are independent of $v_\textrm{bg}^2$ at leading order. 
Even in cases when such a hierarchy in speeds is not respected, we show that one can construct observables leveraging matter-antimatter comparisons where the observable is insensitive to the choice of $v_\textrm{bg}^2$ at leading order, which allows us to avoid potential ambiguities in interpretation related to the specific details of the laboratory reference frame.

Solving the classical equation of motion for $\phi$ in the presence of interaction (\ref{LIV_scalar_interaction}), $\square \phi + m_\phi^2 \phi = - g_\textrm{s} \bar{\psi} (1 - \gamma^0) \psi$, where $m_\phi$ is the scalar mass and the subscript ``s'' denotes the source fermion, we find the following scalar fields sourced by an antimatter and matter particle, respectively, for a non-relativistic pointlike source: 
\begin{equation}
\label{antimatter_source_field}
\phi_\textrm{A} (r) \approx -2 g_\textrm{s} \phi_0  \, , 
\end{equation}
\begin{equation}
\label{matter_source_field}
\phi_\textrm{M} (r) \approx + \frac{g_\textrm{s}}{2 m_\textrm{s}^2}  \left[ \phi_0 \v{p}_\textrm{s}^2 - i \left( \v{\nabla} \phi_0 \right) \cdot \v{p}_\textrm{s} + \v{\sigma}_\textrm{s} \cdot \left( \v{\nabla} \phi_0 \right) \times \v{p}_\textrm{s} \right]  \, , 
\end{equation}
where we have introduced the shorthand $\phi_0 = \exp(-m_\phi r)/(4 \pi r)$. 

To determine the effect of these sourced fields on probe particles, we study the Dirac equation for $\psi$ in the presence of interaction (\ref{LIV_scalar_interaction}): 
\begin{equation}
\label{Dirac_equation_modified}
\left[i \gamma^\mu \partial_\mu - m_\textrm{p} - g_\textrm{p} \left( 1 - \gamma^0 \right) \phi \left( r \right) \right] \psi = 0  \, , 
\end{equation}
where we have introduced the subscript ``p'' to denote the probe fermion. 
The leading interaction of a probe antimatter particle with $\phi$ arises at $\mathcal{O}(1/c^0)$: 
\begin{equation}
\label{antimatter_probe_interaction}
H_\textrm{int}^\textrm{A} \approx + 2 g_\textrm{p} \phi  \, , 
\end{equation}
while the leading interaction of a probe matter particle with $\phi$ arises at $\mathcal{O}(1/c^2)$: 
\begin{equation}
\label{matter_probe_interaction}
H_\textrm{int}^\textrm{M} \approx - \frac{g_\textrm{p}}{2 m_\textrm{p}^2} \left[ \phi \, \v{p}_\textrm{p}^2 - i \left( \v{\nabla} \phi \right) \cdot \v{p}_\textrm{p} + \v{\sigma}_\textrm{p} \cdot \left( \v{\nabla} \phi \right) \times \v{p}_\textrm{p} \right]  \, . 
\end{equation}
Equations (\ref{antimatter_probe_interaction}) and (\ref{matter_probe_interaction}), combined with Eqs.~(\ref{antimatter_source_field}) and (\ref{matter_source_field}), determine the non-relativistic interaction potentials between matter and antimatter particles at leading order (see Appendix~\ref{Appendix:B}).

\textbf{Intra-atomic and intra-nuclear forces} --- 
We begin by studying the phenomenology of interaction (\ref{LIV_scalar_interaction}) associated with the exchange of a virtual scalar $\phi$ within atoms and nuclei. 
We focus on simple two-body systems, where high precision in both experiment and theory is possible. 
Recent work \cite{Jaeckel_2026_Lorentz} has considered operators similar to (\ref{LIV_scalar_interaction}) for various intervals in hydrogen and anti-hydrogen. 
Here we focus on the $1s-2s$ intervals in hydrogen (H), anti-hydrogen ($\bar{\textrm{H}}$), muonium (Mu) and positronium (Ps), as well as the binding energy of deuteron (D). 
A summary of our calculations of the relevant energy shifts is presented in Appendix~\ref{Appendix:C}. 
A summary of our limits and the raw data used to infer them is given in Table~\ref{Tab:LIV_scalar_short-range}.

\begin{table*}[b]
\centering
\caption{ \normalsize 
Summary of $2\sigma$ limits on combinations of coupling parameters defined in interaction (\ref{LIV_scalar_interaction}) from intra-atomic or intra-nuclear forces. 
We have also summarised the experimental and theoretical data used in deriving these limits. 
}
\label{Tab:LIV_scalar_short-range}
\begin{tabular}{ c|c|c|c|c|c|c }%
Atom & $\nu_{1s-2s}^\textrm{expt}(\bar{\textrm{H}})$/kHz & $\nu_{1s-2s}^\textrm{expt}(\textrm{H})$/kHz & $\left| g_e g_p \right|$ limit & $m_\phi$ range & $\left| g_e g_p \right| (m_e \alpha / m_\phi)^2$ limit & $m_\phi$ range  \\ 
\hline 

$\bar{\textrm{H}}$ -- H & $2,466,061,103,079.4(5.4)$ \cite{ALPHA_2018_Hbar_expt} & $2,466,061,103,080.3(0.6)$ \cite{ALPHA_2018_Hbar_expt} & $5.5 \times 10^{-14}$ & $\ll m_e \alpha$ & $1.2 \times 10^{-14}$ & $\gg m_e \alpha$ \\ 
\multicolumn{1}{c}{} & \multicolumn{1}{c}{} & \multicolumn{1}{c}{} & \multicolumn{1}{c}{} & \multicolumn{1}{c}{} & \multicolumn{1}{c}{} & \multicolumn{1}{c}{}  \\
\end{tabular}

\begin{tabular}{ c|c|c|c|c|c|c|c }%
Atom & Coupling & $\nu_{1s-2s}^\textrm{expt}$/MHz & $\nu_{1s-2s}^\textrm{theor}$/MHz & $\left| g_e g_l \right|$ limit & $m_\phi$ range & $\left| g_e g_l \right| (m_e \alpha / m_\phi)^2$ limit & $m_\phi$ range  \\ 
\hline 

Mu & $e - \mu$ & $2,455,528,941.0(9.8)$ \cite{Muonium_1s-2s_2000_expt} & 2,455,528,935.2(1.4) \cite{Muonium_1s-2s_2023_theor} & $8.3 \times 10^{-6}$ & $\ll m_e \alpha$ & $1.9 \times 10^{-6}$ & $\gg m_e \alpha$ \\ \hline
Ps & $e - e$ & $1,233,607,216.4(3.2)$ \cite{Positronium_1s-2s_1993_expt} & $1,233,607,222.18(58)$ \cite{Positronium_1s-2s_1999_theor} & $8.0 \times 10^{-6}$ & $\ll m_e \alpha / 2$ & $7.4 \times 10^{-6}$ & $\gg m_e \alpha / 2$ \\ 
\multicolumn{1}{c}{} & \multicolumn{1}{c}{} & \multicolumn{1}{c}{} & \multicolumn{1}{c}{} & \multicolumn{1}{c}{} & \multicolumn{1}{c}{} & \multicolumn{1}{c}{} & \multicolumn{1}{c}{}  \\
\end{tabular}

\begin{tabular}{ c|c|c|c|c|c|c }%
Nucleus & $E_\textrm{bind}^\textrm{expt}$/MeV & $E_\textrm{bind}^\textrm{theor}$/MeV & $\left| g_p g_n \right|$ limit & $m_\phi$ range & $\left| g_p g_n \right| (200 \, \textrm{MeV} / m_\phi)^2$ limit & $m_\phi$ range   \\ 
\hline 
D & $2.2,245,663(4)$ \cite{Deuteron_binding_energy_1999_expt} & $2.22,457(1)$ \cite{Deuteron_binding_energy_2003_theor,Deuteron_binding_energy_2015_theor} & $1.8 \times 10^{-2}$ & $\ll 100 \, \textrm{MeV}$ & $2.4 \times 10^{-2}$ & $\gg 100 \, \textrm{MeV}$ \\ 
\multicolumn{1}{c}{} & \multicolumn{1}{c}{} & \multicolumn{1}{c}{} & \multicolumn{1}{c}{} & \multicolumn{1}{c}{} & \multicolumn{1}{c}{} & \multicolumn{1}{c}{}  \\
\end{tabular}

\end{table*}

\begin{figure*}[t!]
\centering
\includegraphics[width=8.5cm]{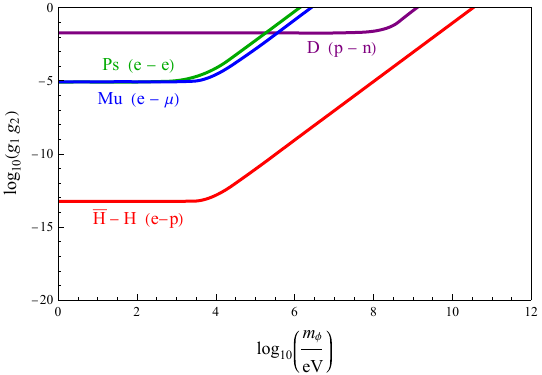}
\caption{ \normalsize (Colour online) 
Bounds ($2\sigma$) on products of dimensionless coupling constants appearing in interaction (\ref{LIV_scalar_interaction}), derived from the $1s-2s$ interval in $\bar{\textrm{H}} - \textrm{H}$ (red), $1s-2s$ interval in muonium (blue), $1s-2s$ interval in positronium (green), and binding energy of deuteron (purple). 
See the main text and Table \ref{Tab:LIV_scalar_short-range} for more details. 
}
\label{Fig:LIV_scalar_short-range}
\end{figure*}

The $1s-2s$ interval in H has been measured with an experimental precision of $\approx 10 \, \textrm{Hz}$ \cite{MPIQ_2011_H_expt}, while the $1s-2s$ interval in $\bar{\textrm{H}}$ has been measured with a precision of $\approx 5 \, \textrm{kHz}$ \cite{ALPHA_2018_Hbar_expt,ALPHA_2025_Hbar_expt}. 
On the other hand, there is a discrepancy in the prediction of the H $1s-2s$ interval at the level of $\sim 1 \, \textrm{MHz}$ depending on whether one uses muonic or electronic atom data for determining the proton radius \cite{CODATA2018,CODATA2022}. 
Additionally, the predicted energy shifts in H depend sensitively on the choice of $v_\textrm{bg}$, see Eq.~(\ref{hydrogen_energy_shift}). 
One can circumvent both of these issues by directly comparing the experimentally measured $1s-2s$ intervals in H and $\bar{\textrm{H}}$ and noting that the SM predicts that these intervals should be identical if the CPT symmetry is respected. 
Using the measured $1s-2s$ interval in $\bar{\textrm{H}}$ and the extrapolated value of the $1s-2s$ interval in H under analogous experimental conditions \cite{ALPHA_2018_Hbar_expt}, combined with the energy shifts in Eqs.~(\ref{anti-hydrogen_energy_shift}) and (\ref{hydrogen_energy_shift}), we place limits on $g_e g_p$ as shown by the solid red line in Fig.~\ref{Fig:LIV_scalar_short-range}. 
We remark that these bounds are dominated by the energy shifts in $\bar{\textrm{H}}$, Eq.~(\ref{anti-hydrogen_energy_shift}), and hence are insensitive to the value of $v_\textrm{bg}^2 \ll 1$ at leading order.

The predicted energy shifts in the leptonic atoms muonium and positronium are insensitive to the assumed choice of $v_\textrm{bg}$ for $v_\textrm{bg} \ll 1$, see Eq.~(\ref{leptonic_atoms_energy_shift}). 
Comparing the measured \cite{Muonium_1s-2s_2000_expt} and predicted \cite{Muonium_1s-2s_2023_theor} values of the muonium $1s-2s$ transition frequency leads to limits on $g_e g_\mu$ shown by the solid blue line in Fig.~\ref{Fig:LIV_scalar_short-range}. 
Likewise, comparing the measured \cite{Positronium_1s-2s_1993_expt} and predicted \cite{Positronium_1s-2s_1999_theor} values of the positronium $1s-2s$ transition frequency leads to limits on $g_e g_e$ shown by the solid green line in Fig.~\ref{Fig:LIV_scalar_short-range}. 
The predicted correction to the deuteron binding energy is insensitive to the choice of $v_\textrm{bg}$ for $v_\textrm{bg} \ll 0.2$. 
Comparing the measured \cite{Deuteron_binding_energy_1999_expt} and predicted \cite{Deuteron_binding_energy_2003_theor,Deuteron_binding_energy_2015_theor} values of the deuteron binding energy, and using Eq.~(\ref{deuteron_binding_energy_shift}) for the predicted shift in binding energy, we find limits on $g_p g_n$ shown by the solid purple line in Fig.~\ref{Fig:LIV_scalar_short-range}.

\textbf{Macroscopic-scale forces} --- 
We now consider the phenomenology of forces between macroscopic bodies in the presence of interaction (\ref{LIV_scalar_interaction}) when the exchanged scalar field $\phi$ changes slowly over the atomic scale. 
In this case, the sourced scalar field and its interactions with matter and antimatter are dominated by the monopolar terms in Eqs.~(\ref{antimatter_source_field},\,\ref{matter_source_field}) and Eqs.~(\ref{antimatter_probe_interaction},\,\ref{matter_probe_interaction}); i.e., the interaction energy between distant bodies falls off as $\propto 1/r$ with the inter-body separation $r$ for $m_\phi r \ll 1$. 
We are specifically interested in the effects of a non-relativistic matter source body on non-relativistic antimatter and matter test bodies, for which the respective interaction energies take the following forms: 
\begin{equation}
\label{matter-antimatter_potential_monopolar}
U^{\textrm{M}-\textrm{A}} (r) \approx + \sum_{\textrm{p},\textrm{s}} \frac{g_\textrm{p} g_\textrm{s}}{m_\textrm{s}^2} \frac{e^{-m_\phi r}}{4 \pi r} \, \left< \v{p}_\textrm{s}^2 \right>  \, , 
\end{equation}
\begin{equation}
\label{matter-matter_potential_monopolar}
U^{\textrm{M}-\textrm{M}'} (r) \approx - \sum_{\textrm{p},\textrm{s}} \frac{g_\textrm{p} g_\textrm{s}}{4 m_\textrm{p}^2 m_\textrm{s}^2} \frac{e^{-m_\phi r}}{4 \pi r} \, \left< \v{p}_\textrm{p}^2 \right> \left< \v{p}_\textrm{s}^2 \right>  \, , 
\end{equation}
where the summations are taken over the different types of source and probe particles and the fermion momenta operators are averaged with respect to the fermion wavefunctions. 
The momentum operators here can be taken to commute with the classical inter-body separation parameter $r$. 
Here we study the effects of interactions (\ref{matter-antimatter_potential_monopolar}) and (\ref{matter-matter_potential_monopolar}) via acceleration of test masses and on atomic spectra. 

\begin{figure*}[t!]
\centering
\includegraphics[width=8.5cm]{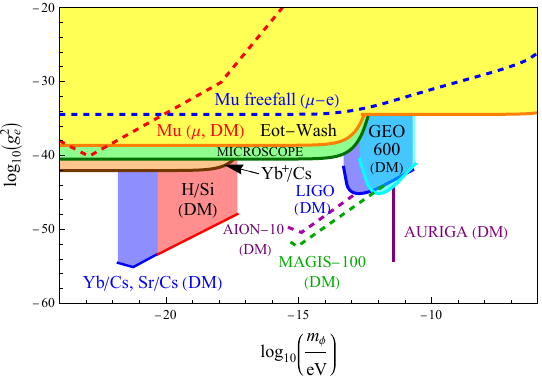}
\includegraphics[width=8.5cm]{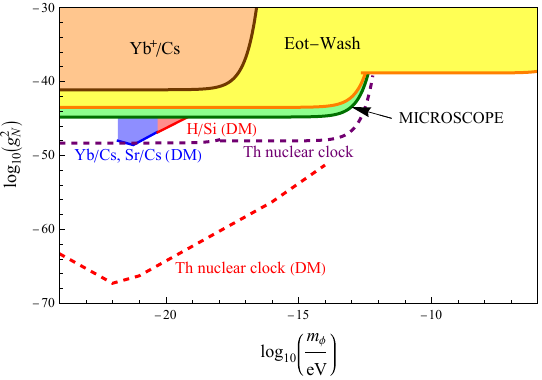}
\caption{ \normalsize (Colour online) 
$2\sigma$ bounds (shaded regions and solid lines) and estimated sensitivities (dashed lines) on the products of dimensionless coupling constants \textbf{(a)} $g_e g_e$ (or $g_e g_\mu$ or $g_\mu g_\mu$ as indicated) and \textbf{(b)} $g_N g_N$ appearing in interaction (\ref{LIV_scalar_interaction}). 
The long-range force limits assume that the elemental composition of Earth is a 1:1:1 ratio of $^{24}$Mg$^{16}$O, $^{28}$Si$^{16}$O$_2$ and $^{56}$Fe by number and that the elemental composition of the Sun is $75\%$ $^1$H and $25\%$ $^4$He by mass. 
We assume $\left< \v{v}_e^2 \right> \sim (\alpha c)^2$ for all atomic electrons and the Fermi-gas-model estimate $\left< \v{v}_N^2 \right> \sim (0.2c)^2$ for all bound nucleons (except in $^1$H, where we assume $\left< \v{v}_p^2 \right> \approx v_\textrm{bg}^2$). 
For the DM-based limits, we assume that the scalar field $\phi$ saturates the local DM density, $\rho_\phi = \rho_\textrm{DM} \approx 0.4~\textrm{GeV/cm}^3$. 
The sensitivities of DM-based experiments to $g^2$ scale as $\propto \rho_\phi$. 
See the main text for more details. 
}
\label{Fig:LIV_scalar_long-range+DM}
\end{figure*}

\emph{Tests of the weak equivalence principle:}
A test body of mass $M_\textrm{test}$ experiences an acceleration $\v{a} = - \v{\nabla}U / M_\textrm{test}$ in the presence of the interactions in Eqs.~(\ref{matter-antimatter_potential_monopolar}) and (\ref{matter-matter_potential_monopolar}). 
Test bodies of different material compositions will generally experience slightly different accelerations in the presence of a common source body, which is commonly parameterised by the E\"{o}tv\"{o}s parameter $\eta_{1,2} = 2(a_1 - a_2) / (a_1 + a_2)$. 
High-precision tests of the weak equivalence principle (WEP) have been performed using Earth as a source body employing ground-based Be$-$Ti and Be$-$Al test mass pairs \cite{EotWash_2008,EotWash_2012} and a space-based Pt$-$Ti test mass pair \cite{MICROSCOPE_2022}, as well as a shorter-range test in the laboratory using a Cu$-$Pb test mass pair in the presence of a uranium source body \cite{RotWash_1999}. 
We place bounds on the electron-electron and nucleon-nucleon coupling parameters in Eq.~(\ref{LIV_scalar_interaction}) using these existing data and present them as the coloured regions in Fig.~\ref{Fig:LIV_scalar_long-range+DM}. 
The ground-based E\"{o}t-Wash bounds correspond to the yellow regions, while the space-based MICROSCOPE bounds are indicated by the green regions. 
Antimatter test masses offer the advantage of a signal that is parametrically suppressed only as $\mathcal{O}(v/c)^2$ versus the $\mathcal{O}(v/c)^4$ scaling for matter test bodies, compare Eqs.~(\ref{matter-antimatter_potential_monopolar}) and (\ref{matter-matter_potential_monopolar}). 
We estimate the expected sensitivity of the ongoing LEMING experiment at PSI \cite{LEMING_2021} to the electron-muon coupling $g_e g_\mu$ based on the target precision of $\Delta g / g \sim 0.1$ using free-falling muonium test atoms and present our projection as the dashed blue curve in Fig.~\ref{Fig:LIV_scalar_long-range+DM}(a).

\emph{Atomic and nuclear spectroscopy:}
When the exchanged scalar field $\phi$ changes slowly over the atomic scale, the effect of the matter-sourced scalar field $\phi \approx + g_\textrm{s} \phi_0 \v{p}_\textrm{s}^2 / (2 m_\textrm{s}^2)$ in Eq.~(\ref{matter_source_field}) on atomic and nuclear energy levels can be inferred from the Dirac equation (\ref{Dirac_equation_modified}) by absorbing the interaction terms into redefinitions of the relevant fermion masses and energies according to: 
\begin{align}
\label{mass_redefinition}
m \to m + g_\textrm{p} \phi \, , \\ 
E \to E \pm g_\textrm{p} \phi \, , 
\label{energy_redefinition}
\end{align}
where the positive (negative) sign in the energy shift is for a matter (antimatter) probe, while the sign of the mass shift is the same for matter and antimatter. 
In this limiting case, the energy shift in Eq.~(\ref{energy_redefinition}) is identical for all atomic electrons or nucleons, hence energy level differences are unaffected by the timelike component of interaction (\ref{LIV_scalar_interaction}). 
This is analogous to a homogeneous electrostatic potential in electromagnetism, which does not affect energy level differences in atoms and nuclei. 

On the other hand, the mass shift in Eq.~(\ref{mass_redefinition}) does affect energy differences between atomic or nuclear levels. 
For example, the comparison of a microwave atomic transition based on a hyperfine interval versus an optical atomic transition based on a regular electronic transition has $\delta (\nu_\textrm{hfs}/\nu_\textrm{opt}) / (\nu_\textrm{hfs}/\nu_\textrm{opt}) \approx \delta m_e / m_e - \delta m_N / m_N$ and thus the observable signal is only parametrically suppressed as $\mathcal{O}(v/c)^2$ versus the $\mathcal{O}(v/c)^4$ scaling in matter-based tests of the WEP (see above). 
High-precision Yb$^+$(E3)/Cs clock comparisons have been used to place constraints on the variation of the electron-to-proton mass ratio correlated with Earth's elliptical orbit about the Sun \cite{PTB_clocks_Sun_2021}. 
Using these data, we derive bounds on $g_e g_e$ and $g_N g_N$ and present them as the orange region in Fig.~\ref{Fig:LIV_scalar_long-range+DM}. 

A significant boost in sensitivity to the nucleon-nucleon coupling may be achieved via the use of nuclear clock spectroscopy based on the low-energy isomeric transition in $^{229}$Th. 
The sensitivity of this transition to variations of the nucleon mass can estimated following the approach in Ref.~\cite{Flambaum_Th_2006} and using the updated transition frequency $\nu_\textrm{Th} \approx 8.4 \, \textrm{eV}$ \cite{Th_clock_2024_expt}, which gives the sensitivity $\delta \nu_\textrm{Th} / \nu_\textrm{Th} \sim + 2 \times 10^5 \, \delta m_N / m_N$. 
We estimate the sensitivity of Th nuclear clock experiments to $g_N g_N$ assuming a fractional frequency accuracy at the $10^{-18}$ level, which is comparable to state-of-the-art optical atomic clocks, and present our projection as the dashed purple curve in Fig.~\ref{Fig:LIV_scalar_long-range+DM}(b). 
Our sensitivity estimate for the scalar masses $m_\phi \lesssim 10^{-18} \, \textrm{eV}$ assumes a ground-based Th clock as Earth orbits the Sun in an elliptical orbit, while for $m_\phi \lesssim 3 \times 10^{-14} \, \textrm{eV}$ we consider a ground-to-space comparison of two Th clocks where one of the clocks is located on a low-earth orbit such as the International Space Station (altitude $\approx 400 \, \textrm{km}$). 


\textbf{Ultralight dark matter} --- 
We now consider the possibility that an ultralight scalar field $\phi$ with interactions (\ref{LIV_scalar_interaction}) constitutes some fraction or the entirety of the DM. 
Ultralight bosons are a good candidate to explain the observed DM. 
Ultralight spinless bosons may be produced non-thermally via the classic ``vacuum misalignment'' mechanism in the early Universe \cite{Wilczek_1983_Axion-cosmo,Sikivie_1983_Axion-cosmo,Fischler_1983_Axion-cosmo} or via the ``thermal vacuum misalignment'' mechanism \cite{Batell_2021_Thermal_misalignment} and can subsequently form a coherently oscillating classical field: 
\begin{equation}
\label{scalar_DM_field}
\phi (t) \approx \phi_\textrm{DM} \cos (\omega_\phi t)  \, , 
\end{equation}
which oscillates at the angular frequency $\omega_\phi \approx m_\phi c^2 / \hbar$. 
The oscillating DM field in Eq.~(\ref{scalar_DM_field}) carries an energy density, averaged over a period of oscillation, of $\left< \rho_\phi \right> \approx m_\phi^2 \phi_\textrm{DM}^2 / 2$. 
In the standard halo model, DM bosons in our local Galactic region are expected to have a root-mean-square speed of $\left< v_\textrm{DM}^2 \right>^{1/2} \sim 10^{-3}c$ relative to the Galactic Center, with a comparable spread in boson speeds. 
The typical spread in the DM boson energies is hence expected to be $\Delta E_\phi / E_\phi \sim \left< v_\textrm{DM}^2 \right> / c^2 \sim 10^{-6}$, implying a coherence time of $\tau_\textrm{coh} \sim 2\pi / \Delta E_\phi \sim 10^6 T_\textrm{osc}$, where $T_\textrm{osc} \approx 2\pi / m_\phi$ is the DM period of oscillation. 

The interactions of antimatter and matter with the DM field $\phi$ are described by Eqs.~(\ref{antimatter_probe_interaction}) and (\ref{matter_probe_interaction}), respectively, and give rise to accelerations on test masses $\v{a} = - \v{\nabla}H_\textrm{int} / M_\textrm{test}$ due to gradients in $\phi$, $|\v{\nabla}\phi| \sim m_\phi v_\textrm{DM} \phi_\textrm{DM}$. 
These accelerations are parametrically suppressed as $\mathcal{O}(v_\textrm{p}^2 v_\textrm{DM}/c^3)$ for matter probes, while antimatter probes are only suppressed as $\mathcal{O}(v_\textrm{DM}/c)$. 
Here we focus instead on the effects of the scalar DM field on the energy levels of atoms and nuclei where such velocity suppression factors are absent altogether for matter and antimatter systems. 
For the relevant DM masses $m_\phi \lesssim 1 \, \textrm{eV}$ corresponding to wavelengths $\lambda_\textrm{DM} \sim 1/ (m_\phi v_\textrm{DM})$ much longer than the atomic scale, the earlier equations (\ref{mass_redefinition}) and (\ref{energy_redefinition}) apply. 
In particular, the mass shifts due to the oscillating DM field (\ref{scalar_DM_field}) take the form: 
\begin{equation}
\label{mass_oscillations}
\frac{\Delta m_\psi}{m_\psi} \approx + \frac{g_\psi \phi_\textrm{DM} \cos \left(m_\phi t \right)}{m_\psi}  \, , 
\end{equation}
for both matter and antimatter fermions. 

Remarkably, in spite of the marked qualitative differences in $\phi$-mediated forces due to (\ref{LIV_scalar_interaction}) compared with standard Yukawa-type forces, the effects of ultralight DM on atomic transition intervals and inter-atomic spacings due to interaction (\ref{LIV_scalar_interaction}) are identical at leading order to DM with the standard Yukawa coupling $\mathcal{L}_\textrm{int} = - g_\psi \phi \bar{\psi} \psi$ \cite{Stadnik_2015_DM-LIFO,Stadnik_2015_DM-VFCs,Stadnik_2016_DM-cavities}.\footnote{We remark that alternative notations are used in the literature for standard Yukawa couplings, including the dimensional energy scale $\Lambda_i = m_i / g_i$ and the dimensionless coupling parameter $d_{m_i} = g_i M_\textrm{Pl} / (\sqrt{4 \pi} m_i)$. See, e.g., Refs.~\cite{Stadnik_2015_DM-LIFO,Hees_2018_EP} for details.} 
This is because atomic energy level intervals and hence atomic structures are unaffected by the timelike component of interaction (\ref{LIV_scalar_interaction}) in the limit of a homogeneous DM field $\phi$. 
We recast existing bounds from Yb/Cs \cite{Yb-Cs_scalar-electron_ULDM_2022} and Sr/Cs \cite{Sr-Cs_scalar-electron_ULDM_2023} atomic clock comparisons, a H/Si clock-cavity comparison \cite{H-Si_scalar-electron_ULDM_2020} and the laser-interferometric gravitational-wave detectors GEO600 \cite{GEO600_scalar-electron_ULDM_2021} and LIGO \cite{LIGO_O3_scalar-electron_ULDM_2024} and present them as coloured regions in Fig.~\ref{Fig:LIV_scalar_long-range+DM}. 
We also recast sensitivity projections for the atom interferometry experiments AION-10 \cite{AION10_scalar-electron_ULDM_2022} and MAGIS-100 \cite{MAGIS100_scalar-electron_ULDM_2021}, as well as existing muonium spectroscopy measurements \cite{Stadnik_2023_DM-muonium,Muonium_HFS_1999_expt}, and present them as dashed lines in Fig.~\ref{Fig:LIV_scalar_long-range+DM}(a). 
Finally, we estimate the sensitivity of future spectroscopy measurements based on a single-ion $^{229}$Th nuclear clock with a fractional frequency instability of $1 \times 10^{-15} \, \tau^{-1/2}$ \cite{Th_single-ion_clock_2012} and present our sensitivity estimate as the dashed red line in Fig.~\ref{Fig:LIV_scalar_long-range+DM}(b). 
For details of the scaling of the DM signal sensitivity with DM mass, see, e.g., Appendix~B of Ref.~\cite{Stadnik_2023_DM-muonium}.

\textbf{Discussion and conclusions} --- 
From Fig.~\ref{Fig:LIV_scalar_long-range+DM}, we can see that the sensitivity of spectroscopy, interferometry and cavity-length-based methods to search for scalar DM with the couplings in (\ref{LIV_scalar_interaction}) is strongly enhanced relative to tests of the WEP (which do not assume that the scalar field comprises any fraction of the DM) compared with the standard Yukawa couplings (compare, e.g., with Figure 12 of Ref.~\cite{Snowmass_scalar_ULDM_2022}). 
The signal in matter-based tests of the WEP is parametrically suppressed as $\mathcal{O}(v/c)^4$ in our Lorentz-violating model, while the aforementioned DM searches have no penalty. 
Furthermore, we also find that clock-based searches for long-range forces in our Lorentz-violating model, being only suppressed as $\mathcal{O}(v/c)^2$, give more stringent bounds than tests of the WEP in the case of the electron coupling and may significantly improve sensitivity to the nucleon coupling using a nuclear clock; this is a different hierarchy compared to the standard Yukawa couplings, see, e.g., Figure 1 in Ref.~\cite{Leefer_2016_EP}. 
Meanwhile, antimatter systems are intrinsically more sensitive to the couplings in (\ref{LIV_scalar_interaction}) over their matter counterparts in spectroscopic searches for intra-atomic forces (see Fig.~\ref{Fig:LIV_scalar_short-range}) and accelerations induced by a scalar field sourced by a massive body or by a scalar field comprising the DM. 
This non-trivial phenomenology presents novel opportunities for a broad range of precision laboratory experiments probing models of new physics that may shed light on the nature of DM and the origins of the matter-antimatter asymmetry.

\textbf{Acknowledgements} --- 
I thank Joerg Jaeckel and Lucas Puetter for helpful discussions.


\begin{appendix}

\section{Complex scalar dark matter model}
\label{Appendix:A}

Here we present a model of a complex scalar DM field $\Phi$ that gives rise to a Lorentz-violating operator structure of the form $\bar{\psi} (1-\gamma^0) \psi$, similar to Eq.~(\ref{LIV_scalar_interaction}). 
Whilst the phenomenology of this complex scalar field model is expected to differ from that of a real scalar field with interaction (\ref{LIV_scalar_interaction}), our intention here is to illustrate that a Lorentz-violating structure like $\bar{\psi} (1-\gamma^0) \psi$ can emerge from a Lorentz covariant theory as a result of the non-relativistic motion of an observer with respect to a preferred reference frame (in this case provided by the Galactic DM medium).


Consider a complex scalar DM field $\Phi \sim \Phi_\textrm{DM} \exp(-i E_\Phi t + i \v{p}_\Phi \cdot \v{r})$ and its complex conjugate $\Phi^\ast \sim \Phi_\textrm{DM} \exp(i E_\Phi t - i \v{p}_\Phi \cdot \v{r})$ with the following interaction that is invariant under an associated global U(1) symmetry: 
\begin{equation}
\label{LIV_complex_scalar_full_interaction}
\mathcal{L}_\textrm{int} \supset -\frac{ig}{2} \Phi \left( \partial_\mu \Phi^\ast \right) \bar{\psi} \gamma^\mu \psi +\frac{ig}{2} \Phi^\ast \left( \partial_\mu \Phi \right) \bar{\psi} \gamma^\mu \psi - g M_\Phi \Phi \Phi^\ast \bar{\psi} \psi  \, . 
\end{equation}
Interaction (\ref{LIV_complex_scalar_full_interaction}) can be rewritten as
\begin{equation}
\label{LIV_complex_scalar_full_interaction_form2}
\mathcal{L}_\textrm{int} \sim g \Phi_\textrm{DM}^2 \left[ E_\Phi \bar{\psi} \gamma^0 \psi - \v{p}_\Phi \cdot \bar{\psi} \v{\gamma} \psi - M_\Phi \bar{\psi} \psi \right]  \, , 
\end{equation}
which in the non-relativistic limit ($E_\Phi \approx M_\Phi$) gives rise to the desired $\bar{\psi} (1-\gamma^0) \psi$ structure. 
If on the other hand the relative sign between the dimension-5 and dimension-6 terms in (\ref{LIV_complex_scalar_full_interaction}) is reversed, the non-relativistic limit instead produces a $\bar{\psi} (1+\gamma^0) \psi$ structure, which projects a Dirac spinor onto its upper component rather than the lower component.

\section{Interaction potentials between matter and antimatter}
\label{Appendix:B}

Here we present the non-relativistic interaction potentials between pointlike matter (M) and antimatter (A) particles at leading order in the coordinate-space representation, which follow from the appropriate combinations of Eqs.~(\ref{antimatter_source_field},\,\ref{matter_source_field}) and Eqs.~(\ref{antimatter_probe_interaction},\,\ref{matter_probe_interaction}): 
\begin{equation}
\label{antimatter-antimatter_potential}
H_\textrm{int}^{\textrm{A}-\textrm{A}'} \approx - \frac{g_\textrm{A} g_{\textrm{A}'} e^{-m_\phi r}}{\pi r} = -4 g_\textrm{A} g_{\textrm{A}'} \phi_0  \, , 
\end{equation}
\begin{equation}
\label{matter-antimatter_potential}
H_\textrm{int}^{\textrm{M}-\textrm{A}} \approx + \frac{g_\textrm{M} g_\textrm{A}}{m_\textrm{M}^2} \left[ \phi_0 \, \v{p}_\textrm{M}^2 - i \left( \v{\nabla} \phi_0 \right) \cdot \v{p}_\textrm{M} + \v{\sigma}_\textrm{M} \cdot \left( \v{\nabla} \phi_0 \right) \times \v{p}_\textrm{M} \right]  \, , 
\end{equation}
\begin{equation}
\label{matter-matter_potential}
H_\textrm{int}^{\textrm{M}-\textrm{M}'} \approx - \frac{g_\textrm{M}}{2 m_\textrm{M}^2} \left[ \phi_{\textrm{M}'} \, \v{p}_\textrm{M}^2 - i \left( \v{\nabla} \phi_{\textrm{M}'} \right) \cdot \v{p}_\textrm{M} + \v{\sigma}_\textrm{M} \cdot \left( \v{\nabla} \phi_{\textrm{M}'} \right) \times \v{p}_\textrm{M} \right]  \, . 
\end{equation}
Here $\phi_0 = \exp(-m_\phi r)/(4 \pi r)$ and $\phi_{\textrm{M}'}$ is given by Eq.~(\ref{matter_source_field}). 
We note that in Eqs.~(\ref{matter-antimatter_potential}) and (\ref{matter-matter_potential}), all momentum operators $\v{p}$ appear to the right of $\phi_0$, which does not commute with the various $\v{p}$. 
We also remark that the classification of ``source'' and ``probe'' particles in Eqs.~(\ref{antimatter_source_field},\,\ref{matter_source_field}) and Eqs.~(\ref{antimatter_probe_interaction},\,\ref{matter_probe_interaction}) is largely semantic when one considers the interaction energy between two bodies, since the interaction energy treats both bodies on an equal footing.

We also present (semi)relativistic expressions for interactions involving matter particle(s): 
\begin{equation}
\label{matter-antimatter_potential_relativistic}
H_\textrm{int}^{\textrm{M}-\textrm{A}} \approx +2 g_\textrm{M} g_{\textrm{A}} \left( 1 - \gamma^0 \right)_\textrm{M} \phi_0  \, , 
\end{equation}
\begin{equation}
\label{matter-matter_potential_relativistic}
H_\textrm{int}^{\textrm{M}-\textrm{M}'} = - g_\textrm{M} g_{\textrm{M}'} \left( 1 - \gamma^0 \right)_\textrm{M} \left( 1 - \gamma^0 \right)_{\textrm{M}'} \phi_0  \, . 
\end{equation}
These expressions are useful in certain cases, e.g., when evaluating energy level shifts in atoms arising from intra-atomic forces.

\section{Calculation of energy shifts in atoms and nuclei}
\label{Appendix:C}
Here we present the results of our calculations of relevant energy shifts in atomic hydrogen (bound state of electron and proton), anti-hydrogen (bound state of positron and anti-proton), muonium (bound state of electron and anti-muon), positronium (bound state of electron and positron), and deuteron (bound state of proton and neutron). 
The relevant atomic wavefunctions for hydrogenlike systems can be found, e.g., in Refs.~\cite{LL3_3rd_Edition,LL4_2nd_Edition}. 

\emph{Hydrogen and anti-hydrogen:} 
Using Eq.~(\ref{antimatter-antimatter_potential}), we find the following energy shift in the $ns$ state of anti-hydrogen: 
\begin{equation}
\label{anti-hydrogen_energy_shift}
\Delta E_{ns} \left( \bar{\textrm{H}} \right)  =  \left< ns \left| H_\textrm{int}^{\textrm{A}-\textrm{A}'} \right| ns \right> \approx \left\{
\begin{aligned}
&- \frac{g_e g_p}{\pi} \frac{m_e \alpha}{n^2}  ~  (m_\phi \ll m_e \alpha) \,  \\ 
&- \frac{4 g_e g_p}{\pi m_\phi^2} \frac{m_e^3 \alpha^3}{n^3}  ~  (m_\phi \gg m_e \alpha) \, , 
\end{aligned}
\right.
\end{equation}
where $n$ denotes the principal quantum number, $m_e$ the electron mass and $\alpha$ the electromagnetic fine-structure constant. 
Using Eqs.~(\ref{matter-matter_potential}) and (\ref{matter-matter_potential_relativistic}), we find the following energy shift in the $ns$ state of hydrogen: 
\begin{equation}
\label{hydrogen_energy_shift}
\Delta E_{ns} \left( \textrm{H} \right)  =  \left< ns \left| H_\textrm{int}^{\textrm{M}-\textrm{M}'} \right| ns \right> \approx \left\{
\begin{aligned}
&- \frac{g_e g_p v_\textrm{bg}^2}{16 \pi} \left[ m_e \alpha^3 \left(\frac{2}{n^3} - \frac{1}{n^4}\right) \right]  ~  (m_\phi \ll m_e \alpha) \,  \\ 
&- \frac{g_e g_p v_\textrm{bg}^2}{4 \pi m_\phi^2} \frac{m_e^3 \alpha^5}{n^3}  ~  (m_\phi \gg m_e \alpha) \, , 
\end{aligned}
\right.
\end{equation}
where we have assumed that $v_\textrm{bg} \sim 10^{-4} - 10^{-3}$ is large compared to the proton's orbital speed about hydrogen's centre of mass, $v_p \sim \alpha m_e / m_p$.

\emph{Muonium and positronium:} 
Using Eqs.~(\ref{matter-antimatter_potential}) and (\ref{matter-antimatter_potential_relativistic}), we find the following energy shift in the $ns$ state of the leptonic atoms muonium ($l = \mu$) and positronium ($l = e$): 
\begin{equation}
\label{leptonic_atoms_energy_shift}
\Delta E_{ns}  =  \left< ns \left| H_\textrm{int}^{\textrm{M}-\textrm{A}} \right| ns \right> \approx \left\{
\begin{aligned}
&+ \frac{g_e g_l}{4 \pi} \left[ m_r \alpha^3 \left(\frac{2}{n^3} - \frac{1}{n^4}\right) \right]  ~  (m_\phi \ll m_r \alpha) \,  \\ 
&+ \frac{g_e g_l}{\pi m_\phi^2} \frac{m_r^3 \alpha^5}{n^3}  ~  (m_\phi \gg m_r \alpha) \, , 
\end{aligned}
\right.
\end{equation}
where the reduced mass of the leptonic atom is $m_r \approx m_e$ in the case of muonium and $m_r = m_e/2$ in the case of positronium. 
In the case of positronium, we have included only the contribution to the inter-fermion interaction potential from the direct t-channel and have omitted the contribution arising from the annihilation s-channel. 
The annihilation contribution is parametrically suppressed compared to the direct contribution by the factor $(v/c)^2 \sim \alpha^2$ in the limiting case $m_\phi \ll m_r \alpha$, though the two contributions may be comparable in size for $m_\phi \gg m_r \alpha$.

\emph{Deuteron}: 
We treat the bound state of deuteron as arising from a simple spherical potential well of depth $U_0$ and radius $R_\textrm{N}$: 
\begin{equation}
\label{deuteron_nuclear_potential}
U_\textrm{nucl}(r) = \left\{
\begin{aligned}
& -U_0 ,  &r < R_\textrm{N} \, , \\ 
& 0 ,  & r > R_\textrm{N} \, . 
\end{aligned}
\right.
\end{equation}
The deuteron bound-state wavefunction is empirically known to be dominated by the $s$-wave contribution, hence we approximate the wavefunction by $\psi(\v{r}) \approx R(r)/\sqrt{4\pi}$. 
Solving the corresponding radial Schr\"{o}dinger equation, $R''(r) + 2R'(r)/r + 2 m_r [E-U(r)] R(r) = 0$, yields the following wavefunction that remains finite as $r \to 0$ and $r \to \infty$: 
\begin{align}
\label{deuteron_radial_wavefunction}
&R_\textrm{I}(r) = \frac{A \sin\left( k r \right)}{r}  \, ,  ~~~~~ r < R_\textrm{N} \, , \\ 
&R_\textrm{II}(r) = \frac{B \exp\left( - \kappa r \right)}{r}  \, ,  ~ r > R_\textrm{N} \, , 
\end{align}
where $k = \sqrt{2 m_r (U_0 - E_B)}$ and $\kappa = \sqrt{2 m_r E_B}$, with $m_r \approx m_N/2 \approx 0.47 \, \textrm{GeV}$ being the deuteron reduced mass and $E_B \approx 2.2 \, \textrm{MeV}$ the binding energy. 
Requiring the continuity of the radial wavefunction $R(r)$ and its first derivative $R'(r)$ at the boundary $r = R_\textrm{N} \approx 2.1 \, \textrm{fm}$, and imposing normalisation of the wavefunction, we determine $A \approx 0.558 \, \textrm{fm}^{-1/2}$, $B \approx 0.875 \, \textrm{fm}^{-1/2}$, and $U_0 \approx 33.5 \, \textrm{MeV}$. 

We estimate the correction to the deuteron binding energy by considering the monopolar term in the interaction potential (\ref{matter-matter_potential}): 
\begin{equation}
\label{deuteron_monopolar_potential}
\Delta E_\textrm{B} \left( D \right) = - \left< D_s \left| H_\textrm{int}^{\textrm{M}-\textrm{M}'} \right| D_s \right> \approx + \frac{g_n g_p}{4 m_N^4} \left< \frac{e^{-m_\phi r}}{4 \pi r} \v{p}^4 \right>  \, , 
\end{equation}
which allows the straightforward application of the Schr\"{o}dinger equation $\v{p}^4 \psi = 4 m_r^2 (U_0 - E_B)^2 \psi$ for $r < R_\textrm{N}$ and $\v{p}^4 \psi = 4 m_r^2 E_B^2 \psi$ for $r > R_\textrm{N}$ in the evaluation of the expectation value. 
In this monopolar approximation, we find the following correction to the deuteron binding energy: 
\begin{equation}
\label{deuteron_binding_energy_shift}
\Delta E_\textrm{B} \left( \textrm{D} \right) \approx \left\{
\begin{aligned}
&+ \frac{g_n g_p}{16 \pi m_N^2} \left[ C_\textrm{I} \left( U_0 - E_B \right)^2 + C_\textrm{II} E_B^2 \right] \approx 1.3 \times 10^{-3} g_n g_p \, \textrm{MeV}  ~  (m_\phi \ll 1/R_\textrm{N}) \,  \\ 
&+ \frac{g_n g_p A^2 k^6}{16 \pi m_N^4 m_\phi^2} \approx \frac{40 g_n g_p \, \textrm{MeV}^3}{m_\phi^2}  ~  (m_\phi \gg 1/R_\textrm{N}) \, , 
\end{aligned}
\right.
\end{equation}
where $C_\textrm{I} = \int_0^{R_\textrm{N}} r [R_\textrm{I}(r)]^2 dr \approx 59.6 \, \textrm{MeV}$ and $C_\textrm{II} = \int_{R_\textrm{N}}^\infty r [R_\textrm{II}(r)]^2 dr \approx 34.9 \, \textrm{MeV}$.

\vspace{20mm}


\end{appendix}


\end{document}